\documentclass[preprint2,numberedappendix,iop]{emulateapj-rtx4}
\usepackage{graphicx}
\usepackage{bm}
\usepackage{natbib}
\newcommand{\approptoinn}[2]{\mathrel{\vcenter{
        \offinterlineskip\halign{\hfil$##$\cr
        #1\propto\cr\noalign{\kern2pt}#1\sim\cr\noalign{\kern-2pt}}}}}

\usepackage{color}

\def\bl{Babcock--Leighton}

\newcommand{\Eq}[1]{Equation~(\ref{#1})}
\newcommand{\Eqs}[2]{equations~(\ref{#1}) and~(\ref{#2})}

\newcommand{\Fig}[1]{Figure~\ref{#1}}
\newcommand{\Tab}[1]{Table~\ref{#1}}

\newcommand{\mps}{m~s$^{-1}$}

\newcommand{\cmss}{cm$^2$~s$^{-1}$}

\newcommand{\Bsat}{B_{\rm sat}}
\newcommand{\taus}{\tau_{\rm s}}
\newcommand{\taup}{\tau_{\rm p}}
\newcommand{\etas}{\eta_{\mathrm{S}}}

\newcommand{\etab}{\eta_{\mathrm{CZ}}}

\graphicspath{{./fig/}{./png/}}

\begin{document}
\title{Dynamo saturation through the latitudinal variation of bipolar magnetic regions in the Sun}
\medskip
\author{Bidya Binay Karak}
\email{karak.phy@iitbhu.ac.in}
\affiliation{Department of Physics, Indian Institute of Technology (Banaras Hindu University), Varanasi, India}
\date{\today}

\begin{abstract}
Observations of the solar magnetic cycle showed that the amplitude of the cycle did not grow all the time in the past.
Thus, there must be a mechanism to halt the growth of the magnetic field in the Sun. We demonstrate 
a recently proposed mechanism
for this under the Babcock--Leighton dynamo framework, which is believed to be the most promising paradigm for the generation of the solar magnetic field at present.
This mechanism 
is based on the observational fact that the stronger solar cycles produce bipolar magnetic regions (BMRs)  
at higher latitudes 
and thus have higher mean latitudes
than the weaker ones. We capture this effect in our three-dimensional Babcock--Leighton solar dynamo model and show that when the toroidal magnetic field tries to grow, it produces BMRs at 
higher latitudes.
The BMRs at higher latitudes generate a less poloidal field, which consequently limits the overall growth of
the magnetic field in our model. 
Thus, our study suggests that the latitudinal variation of BMRs is a potential mechanism for limiting the magnetic field growth in the Sun. 
\end{abstract}
\maketitle


The magnetic cycle in the Sun and other cool late-type stars is believed to be caused 
by a dynamo process operating in the outer convective layers.
In this process, the toroidal component of the magnetic field is largely produced from the poloidal component through
the differential rotation, 
while the poloidal field is recreated back from the toroidal one through the helical convection.
Under certain conditions, this cyclic process continues with an increasing magnetic field, 
if there is no mechanism to halt the amplification.
Although the amplitude of the solar magnetic cycle had cycle-to-cycle-variation in the past, it
did not grow all the time \citep{Uso13}.
Thus there must be a mechanism, the so-called dynamo quenching to halt the overall
growth of the magnetic field in the Sun.
The obvious candidate for this is the Lorentz force of the magnetic field on the flow.
However, due to limited observations, the exact mechanism of the dynamo saturation in the Sun
is still not settled \citep{Cha10,KO11c,Chou14,CDB17}.
The observations of solar differential rotation in the whole convection zone (CZ)
suggest only a tiny variation with the solar cycle \citep{Howe09}.
This variation in the differential rotation alone is unlikely to halt the growth of
the magnetic field in the Sun.
Thus the mechanism of the dynamo saturation might be hidden somewhere in the toroidal $\rightarrow$ poloidal field generation part.

Recent observations \citep{Das10, KO11, Muno13, Priy14, CS15} suggest that the generation of the poloidal field 
in the Sun is primarily through 
the decay and dispersal of tilted BMRs, 
popularly known as the \bl\ process. 
Surface Flux Transport (SFT) models which rely on this process 
remarkably reproduce the magnetic field as observed on the surface of the Sun \citep{Bau04,UH14}.
The dynamo models based on this \bl\ process are also successful in reproducing many 
basic features of the solar magnetic field
and cycle, including short- and long-term variations \citep[e.g.,][]{DC99,CCJ07,Kar10,KC11,CK12,CS17,OK13,KMB18}.
However, all these dynamo models are kinematic and thus we need to invoke a mechanism to
limit the magnetic field growth
in these models.
The usual practice is to include an {\it ad-hoc} nonlinear quenching factor: 
$1/\left[1 + (B/B_0)^2 \right]$ in the poloidal field source \citep{Cha10}.
This quenching implies that when the toroidal magnetic field $B$ exceeds 
the so-called saturation field $B_0$, 
the poloidal field production is reduced.
This type of nonlinear quenching in the \bl\ source, although solves the purpose, has so far no strong physical justification or
 observational support. 

Recent sophisticated \bl\ dynamo models with explicit BMR tilts find stable magnetic cycles again by including
this type of nonlinear quenching in the BMR tilt  \citep{LC17, KM17, KM18}. 
Observations of BMRs for the last two solar cycles find some indication of this tilt quenching \citep{Jha20}; 
also see \citet{Das10}, who found a weak anti-correlation between the cycle-averaged sunspot tilt and the cycle strength. 
Thus this tilt quenching may be a mechanism for the saturation of the solar dynamo.


Another possible mechanism for limiting the growth of the magnetic field in Sun, as highlighted by \citet{J20}, can be the following. 
It is observed 
that the stronger cycles start producing BMRs
at higher latitudes 
and thus have higher mean latitudes 
than the weaker ones 
\citep{Wald55,SWS08, Jiang11, MKB17}. 
The BMRs at higher latitudes are far less efficient
in producing a poloidal field than those at lower latitudes \citep{JCS14}.
On computing the total axial dipole moment at the end of the cycle in an SFT model, 
\citet{J20} showed that this effect acts as a quenching in the growth of the dipole moment 
and thus helps to regulate the solar cycle amplitude. She calls this as the latitudinal quenching. 
This quenching helps to explain the Gnevyshev--Ohl rule.
Further, she showed that the observed tilt quenching, when combined with the latitudinal quenching, leads to 
a saturation in the final dipole moment.
However, 
when there is only latitudinal quenching, 
with the increase of cycle strength, the total dipole moment reduces only 
slightly from the linear dependent. Hence, whether this slight reduction in the dipole moment
is sufficient to stabilize the dynamo growth is not obvious at all. 

We study this problem by performing extensive dynamo simulations of the solar cycle. 
We capture this latitudinal quenching in our novel 3D \bl\ type solar dynamo model.
We find that the BMRs at higher latitudes are far less efficient
in producing a poloidal field than those at lower latitudes---in agreement with the result from SFT model. 
Thus, when a strong cycle produces BMRs
at higher latitudes, it effectively gives less poloidal field. Consequently, the next cycle becomes weak.
This process, stabilize
the growth of the magnetic field in our dynamo model.

\section{Model}
\label{sec:mod}

We perform our study using a recent 3D dynamo model
Surface Transport And Babcock--LEighton (STABLE),
 which is aimed to capture the \bl\ process realistically by utilizing the available surface observations of BMRs and the large-scale flows such as differential rotation and meridional circulation. 
STABLE was primarily developed
 by Mark Miesch \citep{MD14, MT16} and improved by \citet{KM17} to make a close connection of the BMR eruption with observations.
A radial downward magnetic pumping of speed 20~\mps\ is also included in the top $10\%$
of solar radius to mimic the
asymmetric convection. \citet{Ca12} showed that
  this magnetic pumping is essential to make the Babcock--Leighton dynamo models consistent with SFT models.
The pumping helps our model to produce the 11 yr magnetic cycle even at a reasonably high
turbulent diffusivity as inferred from observations \citep[a few times $10^{12}$~\cmss\  in the CZ;][]{CS16},
which was not possible earlier \citep{KC16}.
It also helps the model to recover from grand minima
by reducing the loss of magnetic flux through the surface \citep{KM18}.

As this model does not capture the full dynamics of magnetohydrodynamics convection,
the BMRs do not appear automatically. We have a prescription for this.
First, it computes the strength of the 
azimuthal
field near the base of the CZ
in a hemisphere
\begin{eqnarray}
 \hat{B}(\theta,\phi,t) = \int_{r_a}^{r_b} h(r) B_\phi (r,\theta,\phi,t) dr,
\label{eqspotprodflc}
\end{eqnarray}
where $r_a=0.7 R_\odot$, $r_b=0.715 R_\odot$, and $h(r)=h_0(r-r_a)(r_b-r)$
with $h_0$ being a normalization factor.
The model places a BMR on the surface only when certain conditions are satisfied.
First, $\hat{B}(\theta,\phi,t)$ 
must exceed a critical field strength $B_t(\theta)$.
This critical field depends on the latitude, such that its value exponentially increases
with the latitude in the following way. 
\begin{eqnarray}
 B_t(\theta) = B_{t0} \exp\left[\gamma (\theta-\pi/2) \right], \quad
 \mathrm{for}~~ \theta  > \pi/2 \nonumber\\
 = B_{t0} \exp\left[\gamma (\pi/2-\theta) \right], \quad
 \mathrm{for}~~ \theta  \le \pi/2
 \label{eq:threshold}
\end{eqnarray}
where $\gamma =5$ and $B_{t0} = 2~$kG.
Thus, as latitude increases, the magnetic field has to increase exponentially
to satisfy the condition for BMR eruption. This latitude-dependent threshold plays a crucial role
in capturing the latitudinal quenching in our model. 
Another advantage of using a latitude-dependent threshold is that we do not need to
use any masking function in the \bl\ $\alpha$ (the usual parameterization of the \bl\ process in axisymmetric approximation),
which is needed
in many previous 2D flux transport dynamo models \citep{Dik04,MD14,KC16}.
These masking functions produce very weak variation in the cycle-averaged BMR latitude with the cycle strength---which is in 
contradiction to observations.
There are some possible tachocline instabilities operating in the CZ \citep{PM07,Dik09}
which destabilize the toroidal field to prevent the BMR formation at high latitudes in the Sun.
Particularly, \citet{Kit20} showed that the threshold field strength for the onset of the instability 
of a large-scale toroidal field increases with the increase of the latitude, 
and the growth rate of the instability decreases with latitude. 
We note that only in Sets~A and B, the latitude dependent $B_t(\theta)$, as given by \Eq{eq:threshold}
is considered, while in Set~A$^\prime$,
we simply take $B_t(\theta) = B_{t0}$.

\begin{table}
\caption{
Summary of main runs.
}
\begin{center}
\begin{tabular}{lccccccl}
\hline
Run & Duration &$\Phi_0$& $\sigma_\delta$ &$\tilde{B}_{tor}$ &  $\tilde{B}_{r}$ & Period & Magnetic\\
    & (years)  &        &                 &  (kG)            &    (kG)          & (years) & cycle? \\
\hline
A0  & ~300 & 1.5 & $15^\circ$& --- & --- & --- & Decay\\
A1  & ~400 & 2.0 & $15^\circ$& 12.5& 0.10& 11.7& Stable\\
A2  & 1545 & 2.4 & $15^\circ$& 19.2& 0.18& 10.3& Stable\\
A3  & ~824 & 2.4 & $~0^\circ$& 18.5& 0.17& 10.1& Stable\\
A4  & ~200 & 3.2 & $15^\circ$& 39.3& 0.46& ~7.9& Stable\\
A5  & ~189 & 4.8 & $15^\circ$& --- & --- & --- & Grow \\
A5$^\ast$  & ~185 & 4.8 & $15^\circ$& --- & --- & --- & Grow \\
\hline
A$^\prime$0 & ~123 & 6~ & $15^\circ$  & ---&--- & --- & Decay\\
A$^\prime$1 & ~228 & 8~ & $15^\circ$  & ---&--- & --- & Grow \\
A$^\prime$2 & 1017 & 10 & $15^\circ$  & ---&--- & --- & Grow\\
A$^\prime$2$^\ast$ & ~180 & 10 & $15^\circ$  & ---&--- & --- & Grow\\
\hline
B0 &~100 & 16& $15^\circ$ & --- & --- & ---  & Decay \\
B1 &~404 & 20& $15^\circ$ & 9.2 & 0.08& 11.2 & Stable \\
B2 &~710 & 24& $15^\circ$ & 11.6& 0.12& ~9.5 & Stable \\
B3 &~710 & 48& $15^\circ$ & 17.6& 0.35& ~5.4 & Stable \\
B4 &~231 & 48& $~0^\circ$ & 15.3& 0.27& ~5.8 & Stable \\
B5 &~100 & 64& $15^\circ$ & --- & --- & ---  & Grow  \\
B5$^\ast$
   &~628 & 64& $15^\circ$ & 24.5& 0.66& ~3.8 & Stable \\
\hline
B$^\prime$0 &~100 & 30& $15^\circ$ &  --- & --- & ---  &Decay \\
B$^\prime$1 &~100 & 35& $15^\circ$ &  --- & --- & ---  &Decay\\
B$^\prime$2 &~100 & 40& $15^\circ$ &  --- & --- & ---  &Grow\\
B$^\prime$3 &~100 & 45& $15^\circ$ &  --- & --- & ---  &Grow\\
\hline
\end{tabular}
\tablecomments{
The second column shows the lengths of the simulations,
$\sigma_\delta$ is the standard deviation of the Gaussian tilt scatter around Joy's law,
$\tilde{B}_{tor}$ and $\tilde{B_{r}}$, respectively denote 
average values of the absolute toroidal and poloidal flux densities over
the entire computational domain.
Run~X$\#^\ast$ is the same as Run~X$\#$, except a tilt quenching is included.
The Set A$^\prime$ is the same as A, except the latitudinal quenching is removed.
The Set B$^\prime$ is the same as B, except pumping is put to zero and diffusivity is reduced.
When the dynamo is decaying or growing, the magnetic field and cycle period are dynamic and thus
we do not print their values.
All simulations are performed with spatial resolutions of $200\times 256\times512$ in $r$, $\theta$, and $\phi$.
}
\end{center}
\label{table1}
\end{table}

\begin{figure*}
\centering
\includegraphics[scale=0.95]{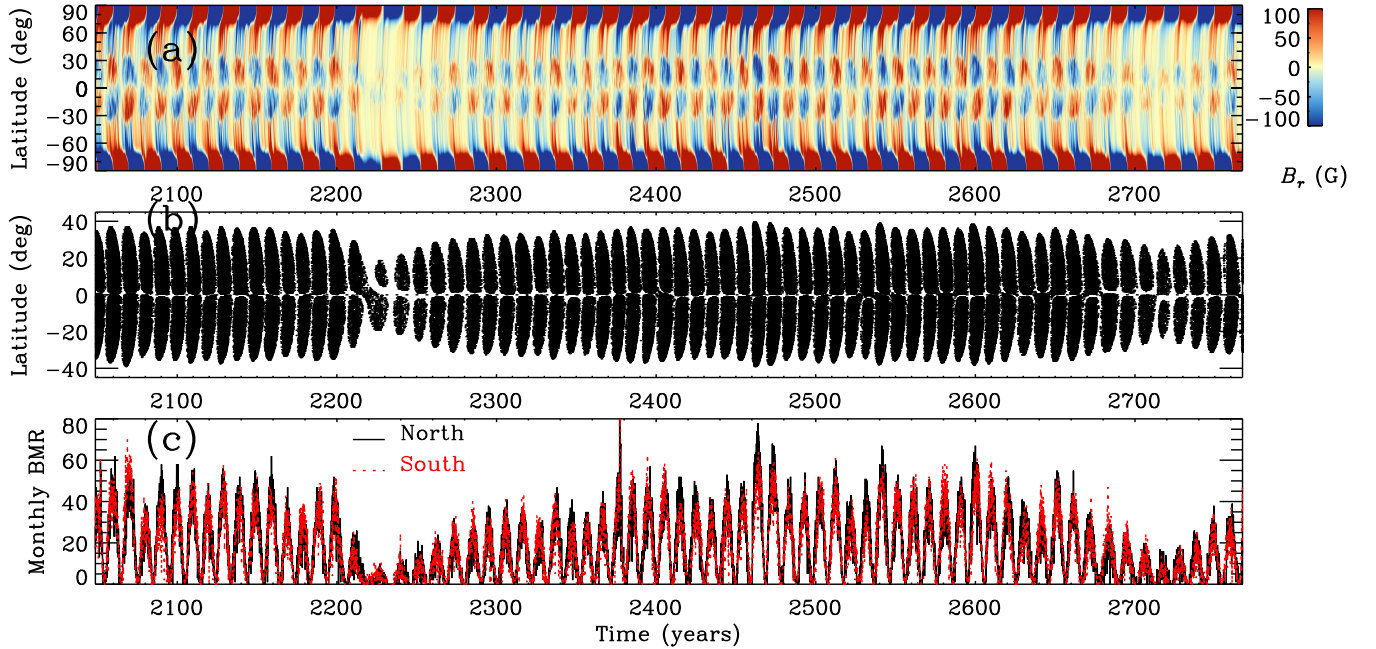}
\caption{
Temporal variations of (a) the azimuthal-averaged surface radial field, 
(b) latitudes of BMRs, and (c) the monthly numbers of BMRs from Run~A2.
}
\label{fig:bflyA2}
\end{figure*}
\begin{figure}
\centering
\includegraphics[scale=0.45]{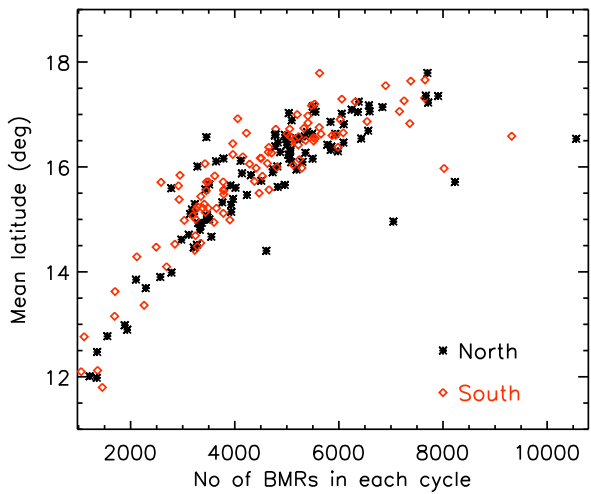}
\caption{
Scatter plot between the mean latitudes of BMRs and the total BMR numbers both computed 
in each cycle.
Red/black are obtained from northern/southern hemispheric cycles.}
\label{fig:SSN_lat}
\end{figure}

Our model produces the first BMR, when $\hat{B}(\theta,\phi) > B_t(\theta)$. Then after a time $dt$
since the previous BMR eruption, the model produces the next BMR only when two conditions,
$\hat{B}(\theta,\phi) > B_t(\theta)$ and $dt \ge \Delta$ are satisfied.
Here $\Delta$ follows a log-normal distribution that is obtained
by fitting the time delay between the observed sunspots:
\begin{eqnarray}
P(\Delta) = \frac{1}{ \sigma_{\rm d} \Delta \sqrt{2\pi} } \exp\left[- \frac{(\ln\Delta - \mu_{\rm d})^2}{2\sigma_{\rm d}^2} \right],
\label{eq:delaypdf}
\end{eqnarray}
where
$\sigma_{\rm d}^2 = (2/3) \left[ \ln\taus - \ln\taup \right] $ and
 $\mu_{\rm d} = \sigma_{\rm d}^2 + \ln\taup$.
In Sets~B and B$^\prime$, we take $\taup=0.8$~days and $\taus=1.9$~days, as derived from the group sunspot data during solar maxima.
However, in Sets~A and A$^\prime$, we consider,
\begin{eqnarray}
\taup = \frac{ 2.2 ~ \mathrm{days}} { 1 + (B_b^N/ B_\tau)^2 }~ ,
\quad
\taus = \frac{ 20 ~ \mathrm{days}} { 1 +(B_b^N / B_\tau)^2 } ~,
\label{eqtau}
\end{eqnarray}
where $B_b^N$ is the azimuthal-averaged toroidal magnetic field in a thin layer
from $r=0.715 R_\odot$ to $0.73 R_\odot$ around $15^\circ$ latitudes and
 $B_\tau = 400$~G.
Hence the delay distribution changes in response to the toroidal field
at the base of the CZ to allow less frequent BMRs when the toroidal field is weak and vice versa.
We note that the whole process is done independently in each hemisphere
so that no hemispheric symmetry is imposed in the flux emergence.

Once the timing of eruption is decided, other properties of BMR
on the surface are obtained
from observations. The field strength of BMR is set to 3~kG,
while the area is obtained
by using the observed distribution of BMR flux \citep{Mu15}:
\begin{eqnarray}
P(\Phi) = \Phi_0 \frac{1}{ \sigma_\Phi \Phi \sqrt{2\pi} } \exp\left[- \frac{(\ln\Phi - \mu_\Phi)^2}{2\sigma_\Phi^2} \right],
\label{eqflux}
\end{eqnarray}
with $\mu_\Phi=51.2$, and $\sigma_\Phi =0.77$. The factor $\Phi_0$ regulates the strength of the dynamo (or the dynamo number);
see Table~1.

To emphasis our prescription, in Sets~A and A$^\prime$, it is the BMR time delay part through which the toroidal field
is linked to the BMRs. 
The BMR spot flux is taken from \Eq{eqflux}.
However, in Sets~B and B$^\prime$ the toroidal field is linked in a different way.
As mentioned above, in these sets, the delay distribution is kept unchanged, but the (observed) BMR flux distribution is scaled linearly with the toroidal field at the base of the CZ. 
Thus, in 
these sets, 
the BMR spot flux
$\Phi_s = (\hat{B}(\theta_s,\phi_s,t) / \Bsat) \Phi$, where
($\theta_s,\phi_s$) is the location of the BMR, and
$\Phi$ is obtained from \Eq{eqflux}.

For the tilts of BMRs, we consider Joy's law with a Gaussian scatter around it with a given $\sigma_\delta$
inferred from observations \citep{Das10, SK12, MNL14, Wang15, Arlt16, Jha20}. 
For further details of the model, readers are encouraged to go through \citet{KM17}.


\section{Results}
\label{sec:res}
In \Tab{table1}, we enlist some of the key parameters and results of our primary simulations. 
In Set~A, Run~A0 is subcritical, while all other Runs produce dynamo cycles. 
Runs~A1--A4 produce a stable magnetic field. 
In \Fig{fig:bflyA2}, we show the time evolutions of various quantities 
for about 700 years from Run~A2. 
We observe that this simulation produces an overall stable magnetic field 
even without including any explicit nonlinear quenching.
After continuing this simulation for 1545~years, we find 
that the magnetic field overall remains stable.
Other than the stable magnetic field, the simulation produces 
most of the basic features of
the solar cycle, namely, polarity reversals, 
dipole dominated field near minima, 
amplitude variation, north-south asymmetry, and 
mixed-polarity field.
All Runs~A1--A4 show these features. The variation in the magnetic field
in these simulations is due to the scatter in BMR tilt and the randomness in BMR emergences.
We note that in our modeled solar cycle, we do not observe the Gnevyshev-Ohl rule even when we include the tilt quenching
in addition to the latitude quenching. 
Hence, the prediction of \citet{J20} based on the SFT model does not hold in our dynamo model.

Another feature that we observe in this simulation is that the latitudinal extent of the BMRs in each cycle is not constant. 
As seen in \Fig{fig:bflyA2}b, stronger cycles start producing BMRs at
higher latitudes, while weaker ones produce BMRs at lower latitudes.
This feature is nicely seen in \Fig{fig:SSN_lat}. It shows a positive trend between the mean latitude of BMRs and the total number of BMRs in each cycle, consistent with observations 
\citep{Wald55,SWS08, Jiang11, MKB17}.
It is this feature of our model, which gives rise to the latitudinal quenching and stabilizes the magnetic field growth.

\begin{figure}
\centering
\includegraphics[scale=0.42]{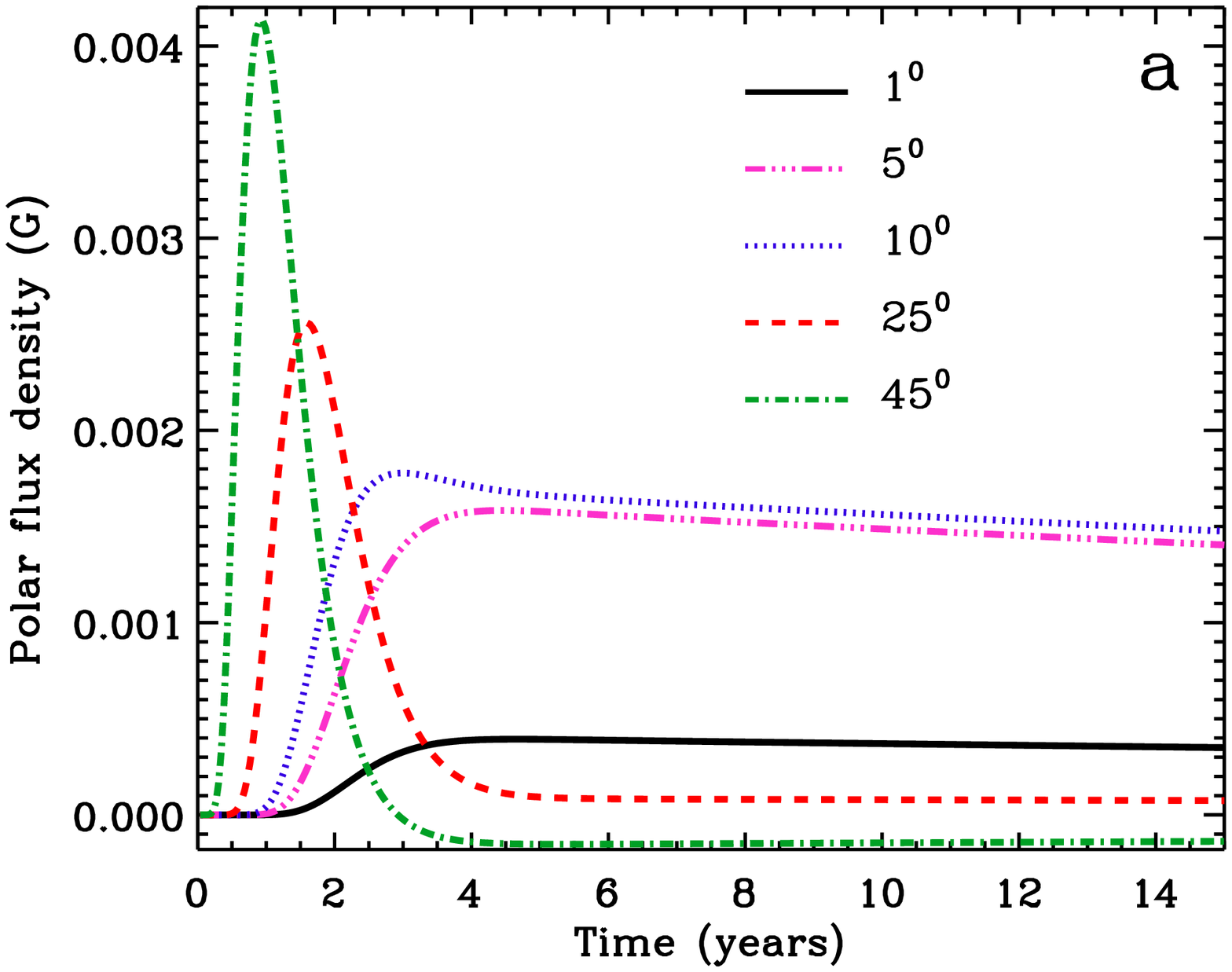}
\includegraphics[scale=0.42]{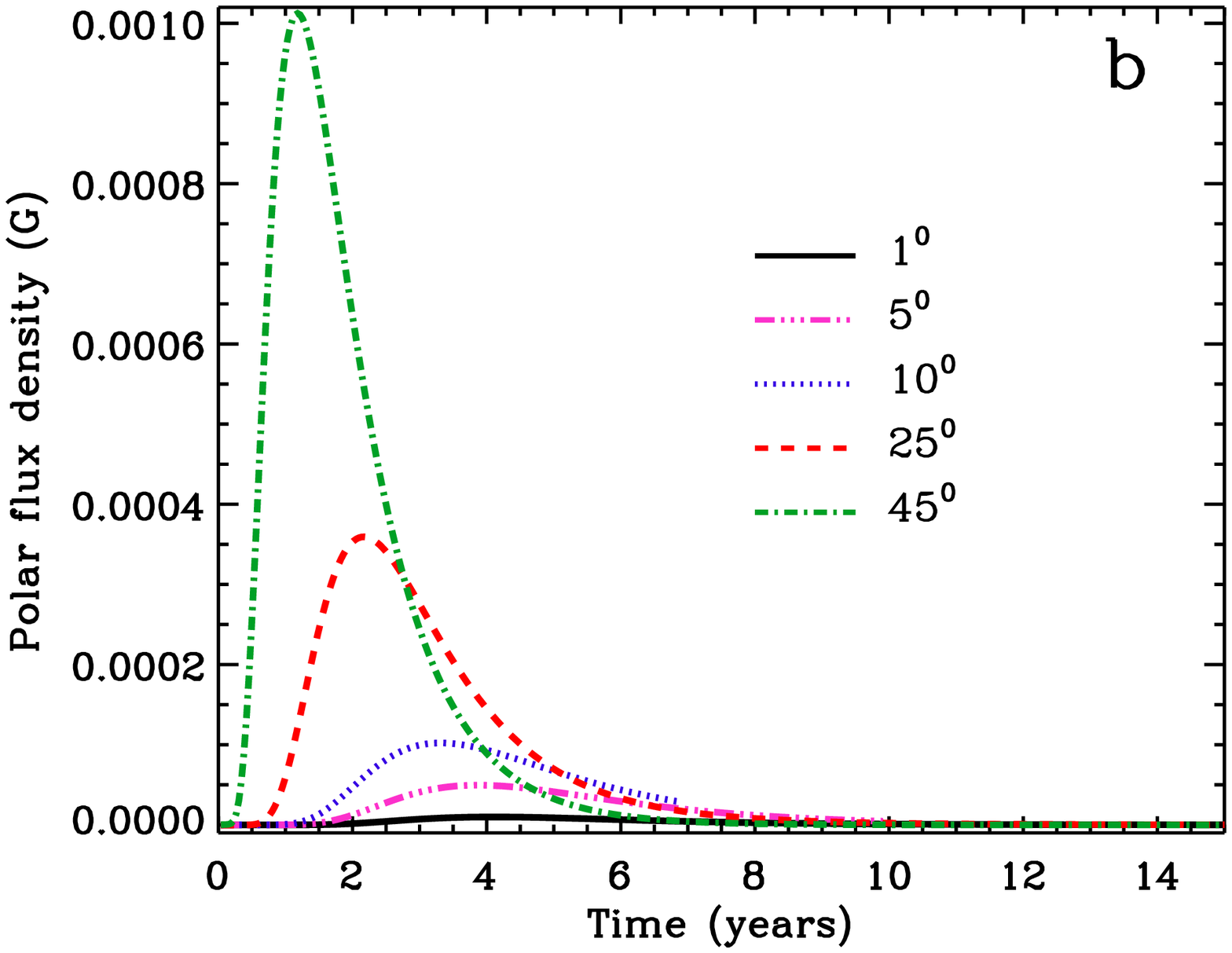}
\caption{
Time evolution of the polar magnetic flux density averaged over $55^\circ$ to pole 
in the northern hemisphere. Solid, 
dash double-dot, dot, dash, and dash-dot lines correspond to simulations
with BMR pair 
at $\pm1^\circ$, $\pm5^\circ$, $\pm10^\circ$, $\pm25^\circ$, and $\pm45^\circ$ latitudes, respectively.
(a) and (b) correspond to simulations with and without magnetic pumping, respectively.
}
\label{fig:twoBMRs}
\end{figure}

To demonstrate how this is happening in our model, we make the following clean experiment. 
We perform four simulations by depositing two identical BMRs at latitudes: $\pm1^\circ$, $\pm 5^\circ$,
$\pm 10^\circ$, $\pm 25^\circ$, and $\pm 45^\circ$. 
We mean, at the beginning of each simulation, we deposit one BMR at a given latitude in the northern hemisphere 
and another one exactly at the same latitude in the southern hemisphere.
The tilt is assigned by Joy's law with no scatter around it.
No other initial magnetic field is given. The time evolutions of the surface radial magnetic flux densities averaged over $55^\circ$ to the pole from these simulations are shown in \Fig{fig:twoBMRs}a. 
We observe that with the increase of BMR latitude, from $1^\circ$ to $10^\circ$, the polar flux increases 
(due to the increase of tilt).  However,
then with the increase of latitude, the polar flux rapidly decreases 
(due to less efficient cancellation of the opposite polarity flux at the equator). 
The BMR pair at $\pm 45^\circ$ gives even little negative polar flux.
Thus in our dynamo simulation, the BMRs at higher latitudes give rise to less polar flux
which is also true in the SFT model \citep{JCS14}.

This behavior does not hold entirely if we do not include the downward magnetic pumping in our model.
As seen in \Fig{fig:twoBMRs}b, the polar field in the early phase from models without pumping behaves similarly 
to that from models with pumping. However, after a few years, the magnetic field decays quickly; \cite{HCM17} also found a similar behavior.
This does not happen when there is pumping. It was already realized in previous studies that a magnetic pumping is needed to make the 
results of dynamo models consistent with SFT models and the observations \citep{Ca12,KM18}.

Thus in our model, due to the inclusion of latitude-dependent threshold for BMR eruption (\Eq{eq:threshold}), 
when the toroidal magnetic field tries to grow in one cycle,
the mean latitude of BMRs increases. As seen above, the BMRs at higher latitudes are far less efficient in generating poloidal field. This effect halts the growth of the magnetic field in our dynamo model.

As expected, when we remove the latitude-dependent threshold for BMR eruption in Set~A$^\prime$,
the dynamo cannot produce a stable magnetic field. This happened in Runs~A$^\prime$1--A$^\prime$2.
However, Run~A$^\prime$2$^\ast$, which is the same as Run~A$^\prime$2 except a tilt quenching of the form $1/[1 + ( \hat{B}(\theta,\phi,t) /\Bsat)^2]$ 
(where the saturation field $\Bsat = 1\times10^5$~G) is included, also fails to 
produce a stable magnetic field.

We have made several simulations by changing some parameters in the model.
Run A3 is the case in which we have switched off the scatter in the BMR tilt
around Joy's law. We again find a stable solution.
However, when $\Phi_0$ is sufficiently above the critical value needed for the dynamo transition, 
the model fails to produce a stable magnetic field. This is
because of an opposing effect arose at large $\Phi_0$.
When $\Phi_0$ is large, the BMR flux distribution (\Eq{eqflux}) is moved to a higher side.
Thus the individual BMR gets more flux, which consequently generates a large poloidal flux. 
When this effect dominates over the reduction of poloidal flux by the latitudinal quenching, 
the model fails to provide a stable magnetic field. This happened in Run~A5.
Interestingly, in this run when we include a tilt quenching, it also fails to 
produce a stable magnetic field (Run~A5$^\ast$ in \Tab{table1}).
Thus when the dynamo is too much supercritical, both the latitudinal and tilt quenchings, even operating together,
fail to produce a stable magnetic field in Set~A.

We note that in Sets~A and A$^\prime$, when a model fails to limit the magnetic field growth 
at large $\Phi_0$, the magnetic field
cannot grow indefinitely because of a numerical means.
At large $\Phi_0$, the strong magnetic field makes the delay distribution for BMR eruptions
narrow (\Eqs{eq:delaypdf}{eqtau}). Meaning, $\Delta$ is less. However,
the actual $\Delta$ cannot be less than the time step of numerical
integration of our differential equations. Hence, when the magnetic
field is sufficiently high to make the delay less than or equal to
the numerical time step, the growth of the
magnetic field is artificially halted.
However, this does not happen when $\Phi_0$ is not too much larger than the critical value for the dynamo transition.
Thus in Runs~A1--A3 the delay never became less than the numerical
time step and the magnetic field growth
is limited by the latitudinal quenching alone.

\begin{figure}
\centering
\includegraphics[scale=0.92]{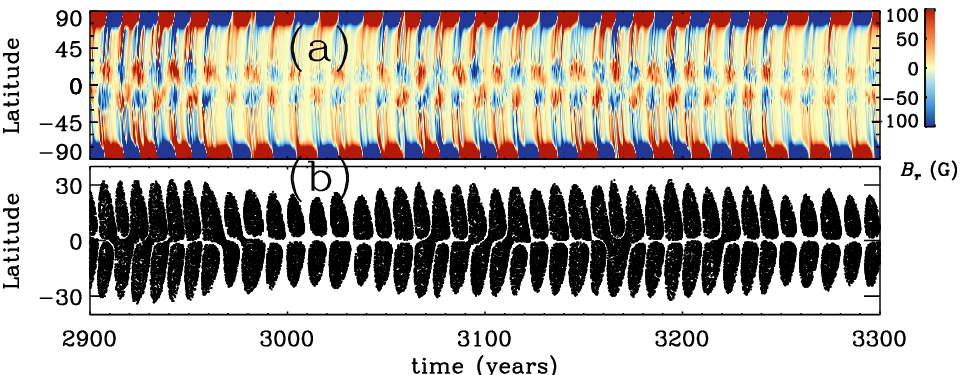}
\caption{The same as \Fig{fig:bflyA2} but obtained from Run~B2.
}
\label{fig:bflyB2}
\end{figure}

This issue is not there in Set~B because in this case, the delay distribution is
kept at the observed value.
Again in Runs B1--B4, we find a stable solution without including any nonlinearly in the model.
Only the latitudinal effect of BMR eruption as discussed above is responsible for the stability of the magnetic field. 
Time evolutions of various quantities from Run~B2 are presented in \Fig{fig:bflyB2}. 
Again we observe a stable magnetic field and the basic features of the observed solar magnetic field. 
However, when we make the dynamo too strong by increasing $\Phi_0$ much above the critical value, 
we examine that
the latitudinal quenching is not able to halt the growth of the magnetic field.
This is seen in Run B5. Nevertheless, when we include the tilt quenching, the model
manages to halt the magnetic field growth in this case; see Run~B5$^\ast$.

Based on the physics presented in \Fig{fig:twoBMRs}, it is expected that if we exclude the magnetic pumping, 
the model fails to produce a stable magnetic cycle
through the latitudinal quenching alone. To demonstrate this, 
we show a few additional simulations Runs~B$^\prime$0--B$^\prime$3.
We note that when we put pumping to zero, the dynamo model fails to produce growing field unless
we reduce the diffusivity in the CZ considerably or increase $\Phi_0$ to a very high value.
It is already known that the flux transport dynamo cannot produce 11-year solar cycle at high
diffusivity ($\sim 10^{12}$~\cmss) \citep{KC12}. Thus we reduce the diffusivity
in this runs by taking $\etab = 5 \times 10^{10}$~cm$^2$~s$^{-1}$, and
$\etas = 1 \times 10^{12}$~cm$^2$~s$^{-1}$.
Other than this change in the diffusivity, the B$^\prime$ Set in \Tab{table1}
is the same as Set~B.
From \Tab{table1} we observe that when $\Phi_0$ is above a certain value, the model produces growing field
but no stable magnetic cycle.

\section{Summary and Conclusions}
We have demonstrated the saturation of the
magnetic field in the kinematic \bl\ type flux transport dynamo
models through the latitudinal quenching as proposed by \citet{J20}.
It is based on the observed fact that the stronger cycles
produce sunspots and BMRs at higher latitudes
than the weaker ones.
The BMRs at higher latitudes are less efficient in
producing the poloidal field. This effect alone halts the growth of the
magnetic field in our dynamo model. However, when the
dynamo is too much supercritical (much above the dynamo transition), the
latitudinal quenching cannot limit the growth of the magnetic
field in our model.
Incidentally, there are some indications that the solar dynamo is not too much supercritical \citep{Met16,KN17}.
Thus we conclude that the latitudinal variation of BMR is a potential candidate for the saturation of the solar dynamo. 



\begin{acknowledgements}
We thank the anonymous referee for making critical comments and suggestions
that helped to improve the presentation.
Financial supports from Department of Science and Technology
(SERB/DST), India through the Ramanujan Fellowship (project no
SB/S2/RJN-017/2018) and ISRO/RESPOND (project no ISRO/RES/2/430/19-20)
are acknowledged. 
The Computational support and the resources provided by PARAM Shivay Facility under the National Supercomputing Mission, Government of India at the Indian Institute of Technology, Varanasi are gratefully acknowledged.
The author thanks Mark Miesch, the developer of the original STABLE code,
for constant help and support when the author was a Jack Eddy Postdoctoral fellow at High Altitude Observatory.
The author further acknowledges the support from the International Space Science Institute (ISSI) Team 474
\end{acknowledgements}

\bibliographystyle{apj}
\bibliography{paper}

\end{document}